\documentclass{webofc}
\usepackage[varg]{txfonts}   
\begin{document}
\title{Relativistic SZ maps and electron gas temperature spectroscopy}
%
%

\author{\firstname{Mathieu} \lastname{Remazeilles}\inst{1}\fnsep\thanks{\email{mathieu.remazeilles@manchester.ac.uk}} 
}

\institute{Jodrell Bank Centre for Astrophysics, Department of Physics and Astronomy, The University of Manchester, Manchester M13 9PL, UK
          }

\abstract{%
  While third-generation CMB experiments have allowed to release the first maps of Compton-$y$ distortion due to thermal Sunyaev-Zeldovich (SZ) effect, next-generation CMB experiments should allow us to map also the electron gas temperature,  $T_{\rm e}$, across the sky through the detection of relativistic corrections to the thermal SZ effect. We discuss about experimental requirements to break the  $y$-$T_{\rm e}$  degeneracy of the observed SZ intensity, and propose a new component separation approach based on moment expansion to disentangle the  $y$  and  $T_{\rm e}$  observables of the relativistic SZ effect while mitigating foregrounds. We show how our approach offers a new spectroscopic view of the clusters not only across frequencies but now also across temperatures. We also show how the relativistic electron temperature power spectrum provides a new cosmological observable which may complement the Compton-$y$ map power spectrum to break some of the parameter degeneracies in future cosmological SZ analyses.
}
\maketitle
\section{Introduction}
\label{sec:intro}

By crossing the hot gas of electrons in galaxy clusters and filaments, cosmic microwave background (CMB) photons get scattered while receiving a boost of energy as the result of inverse Compton scattering. This is known as the thermal Sunyaev-Zeldovich (SZ) effect \cite{sz}, which causes $y$-type spectral distortions to the primary CMB blackbody spectrum prominently in the direction $\hat{n}$ of galaxy clusters:
\begin{equation}\label{eq:tsz}
I_\nu^{\rm SZ}(\hat{n}) \equiv \frac{\Delta I_\nu^{\rm CMB}}{I_\nu^{\rm CMB}} (\hat{n})= g(\nu)\, y(\hat{n})\,.
\end{equation}
The amplitude of the distortion given by the Compton parameter $y$ depends on the integrated pressure profile $P_{\rm e}$ of the electron gas along the line-of-sight as
\begin{equation}\label{eq:y}
y = \frac{\sigma_T}{m_{\rm e}c^2}\int P_{\rm e}(l)dl\,,
\end{equation}
where $m_{\rm e}$ is the electron mass, $c$ is the speed of light, and $\sigma_T$ is the Thomson scattering cross-section. The spectral shape $g(\nu)$ of the distortion across the frequencies $\nu$ is given in the \emph{non-relativistic} limit, i.e. $kT_{\rm e}/m_{\rm e}c^2 \ll 1$, by \cite{sz}
\begin{equation}\label{eq:nrsed}
g(\nu) \equiv f(\nu, T_{\rm e}\simeq 0) = \frac{x^4e^x}{\left(e^x-1\right)^2}\left[x\coth \left(\frac{x}{2}\right) -4 \right]\,,
\end{equation}
where $T_{\rm e}$ is the electron gas temperature, $h$ is the Planck constant, $k$ is the Boltzmann constant, $T_{\rm CMB}=2.7255$\,K is the CMB blackbody temperature, and $x\equiv h\nu / k T_{\rm CMB}$.

\begin{figure*}
\centering
\includegraphics[width=0.5\columnwidth]{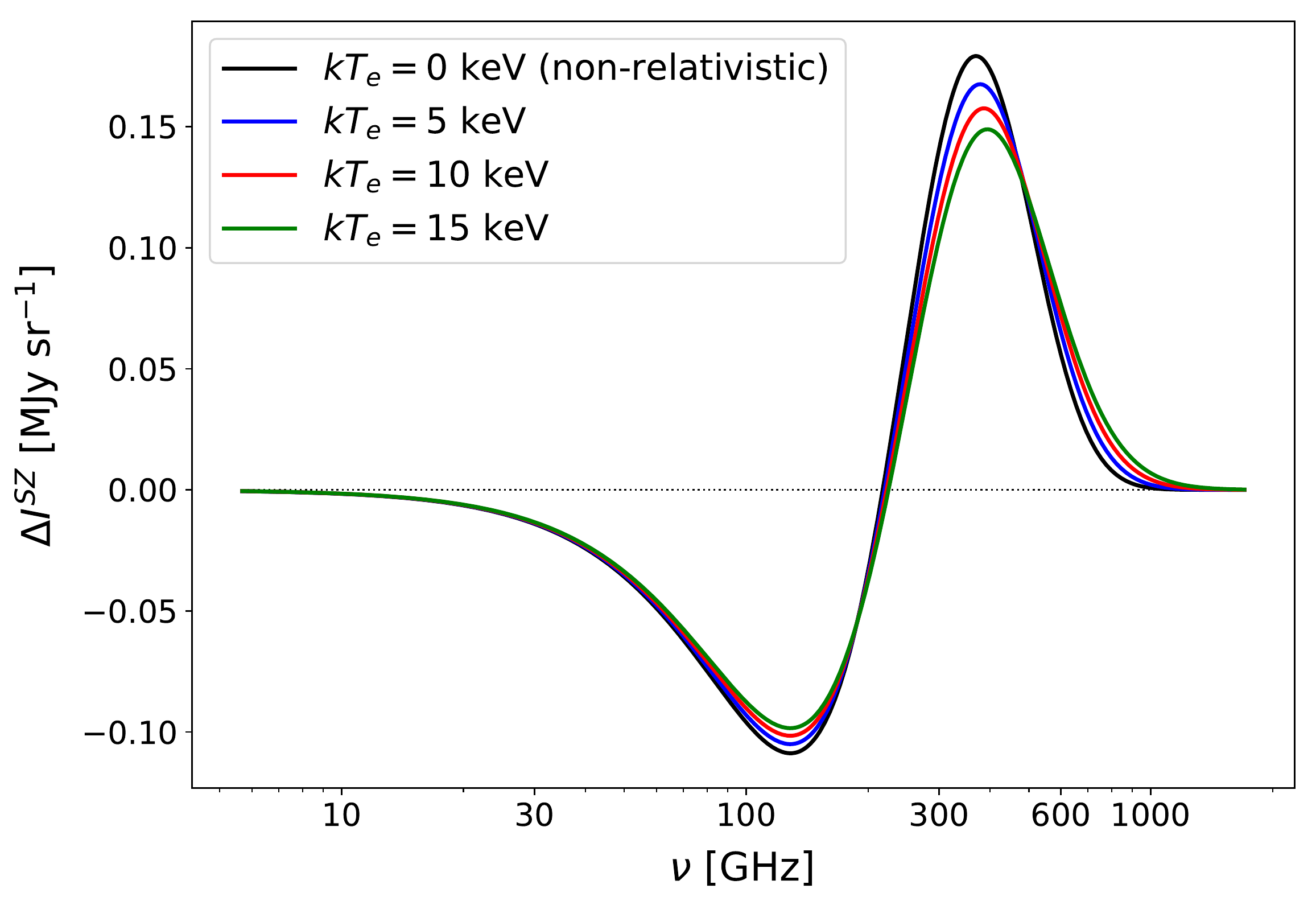}~
\includegraphics[width=0.5\columnwidth]{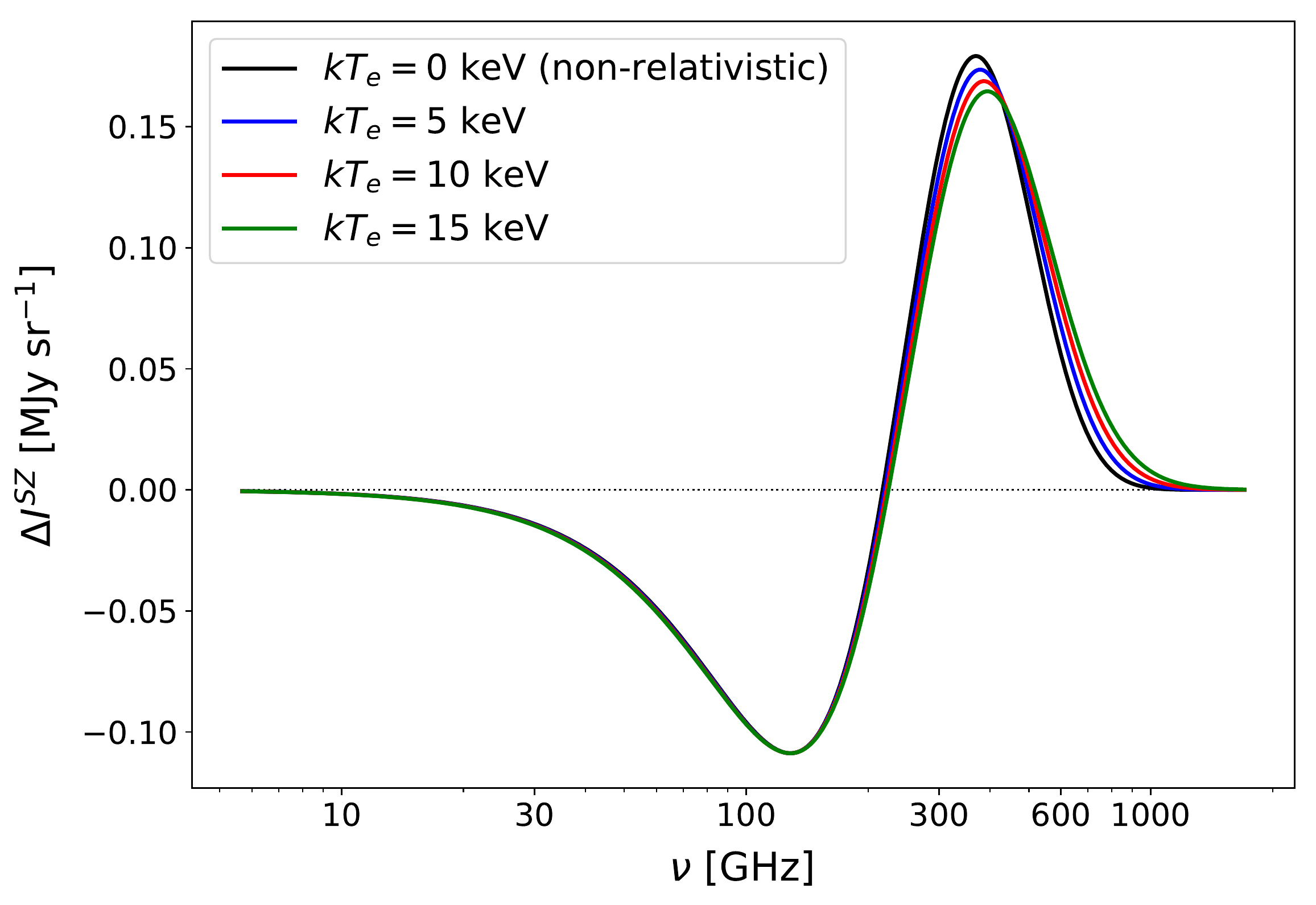}
\caption{\emph{Left}: Variations of the thermal SZ spectrum with the electron temperature $T_{\rm e}$ at fixed $y = 10^{-4}$, using \textsc{SZpack} \cite{Chluba2013}. \emph{Right}: A varying Compton-$y$ parameter is used to rescale the amplitudes of the spectra and make them matching at low frequencies, which allows us to highlight that the distortion of the SZ spectral shape due to relativistic electron temperature corrections actually arises at high frequencies.}
\label{fig:seds}       
\end{figure*}

The peculiar spectral signature Eq.~\eqref{eq:nrsed} of the thermal SZ effect, as illustrated in Fig.~\ref{fig:seds} by the black solid line, with a decrement at low frequencies, a null around $218$\,GHz and an increment at high frequencies, enables to detect galaxy clusters by spectroscopy across frequencies \cite{Planck_sz_2011}. Hence, the thermal SZ effect has now become a routine tool of third-generation CMB experiments to detect thousands of galaxy clusters (e.g. \cite{Planck_sz_2015}) and trace the hot gas in the Universe through the mapping of Compton-$y$ parameter anisotropies across the sky \cite{Planck_ymap_2015}.

However, galaxy clusters are massive,  therefore they are hot objects with typical electron gas temperature of $kT_{\rm e}\simeq 5$\,keV for typical cluster mass of $M_{500} = 3\times 10^{14}\,h^{-1}M_\odot$ \cite{Arnaud2005}. Hence, the thermal velocity of electrons in the gas approach the speed of light, ${v_{\rm e}^{\rm th} = \sqrt{2kT_{\rm e}/m_e}\gtrsim 0.1c}$. Therefore, relativistic electron temperature corrections to the thermal SZ effect become relevant:
\begin{equation}\label{eq:rsz}
I_\nu^{\rm SZ}(\hat{n}) = f(\nu, T_{\rm e}(\hat{n}))\, y(\hat{n})\,.
\end{equation}
The spectral signature $f(\nu, T_{\rm e}(\hat{n}))$ of the \emph{relativistic} SZ (rSZ) effect thus varies depending on the local electron gas temperature in the sky (Fig.~\ref{fig:seds}). 
While at \textit{Planck} sensitivity relativistic temperature corrections are not perceptible on individual clusters, they become statistically significant for a large number of clusters \cite{Hurier2016,Erler2018,Remazeilles2019}. 
With larger sensitivity and larger sets of frequencies, next-generation CMB experiments \cite{Hazumi2020,Hanany2019} should allow us to not only map the thermal SZ Compton parameter $y(\hat{n})$ but also the relativistic electron temperature field $T_{\rm e}(\hat{n})$ across the sky. In this proceeding paper we describe a new component separation approach allowing us to disentangle the $y$ and $T_{\rm e}$ observables of the rSZ effect \cite{Remazeilles2020}, and show how the extracted $T_{\rm e}$ field can complement the Compton-$y$ observable to probe cosmology and astrophysics with galaxy clusters.

\section{Disentangling the $y$ and $T_{\rm e}$ observables of the relativistic SZ effect}
\label{sec:compsep}

As shown in the left panel of Fig.~\ref{fig:seds}, relativistic corrections reduce the overall intensity of the thermal SZ signal. Therefore, for a measured intensity of the signal across the frequencies, the assumption of the non-relativistic limit (black line) of the spectrum to detect the SZ effect from hot galaxy clusters will result in underestimating the Compton-$y$ parameter. 
As shown in \cite{Remazeilles2019}, neglecting relativistic SZ corrections in the \emph{Planck} data analysis may have led to an underestimation of the overall amplitude of the \emph{Planck} $y$-map power spectrum, and thereby the inferred $\sigma_8$ value. Future cosmological analyses will thus have to account for rSZ effects. 

\subsection{Experimental requirements}
\label{subsec:exp}

In the right panel of Fig.~\ref{fig:seds}, we rescaled the various rSZ spectra of different temperatures $T_{\rm e}$ by varying the parameter $y$ so that they match at low frequency. This rescaling allows us to highlight that relativistic electron temperatures actually distort the spectral shape of the thermal SZ signal mostly in the high-frequency range $\gtrsim 300$\,GHz. At low frequencies $< 200$\,GHz, the rSZ effect does not cause any spectral distortion to the signal (right panel of  Fig.~\ref{fig:seds}) but only a change in overall amplitude (left panel of  Fig.~\ref{fig:seds}).  Therefore, there is a $y$-$T_{\rm e}$ degeneracy of the rSZ signal at low frequencies, where a lower amplitude of the measured intensity can either be attributed to lower $y$ or larger $T_{\rm e}$ without any possible distinction. In order to break this spectral degeneracy and disentangle the $y$ and $T_{\rm e}$ observables of the rSZ effect, it is thus recommended to observe the sky at high frequencies $> 200$\,GHz. Fourth-generation CMB satellite concepts, e.g. \emph{LiteBIRD} \cite{Hazumi2020} or \emph{PICO} \cite{Hanany2019}, which are designed to observe the sky from space at frequencies above $300$\,GHz with high sensitivity, are thus best suited for our purpose.

\subsection{Sky map simulations}
\label{subsec:sims}

We generated full-sky map simulations in 21 frequency bands ranging from $21$ to $800$\,GHz using the instrumental specifications of \emph{PICO} \cite{Hanany2019} in order to test the ability of our component separation method to disentangle the $y$ and $T_{\rm e}$ observables of the rSZ effect. Our sky simulations include the relativistic SZ signal in each frequency channel, which we generated from all-sky template maps of $y$ and $T_{\rm e}$ (see figure 3 in \cite{Remazeilles2020}) using the frequency scaling Eq.~\eqref{eq:rsz} computed with \textsc{SZpack} \cite{Chluba2013}. Our simulations also include several foreground emissions: kinetic SZ effect, CMB anisotropies, cosmic infrared background anisotropies, synchrotron, anomalous microwave emission, free-free and thermal dust (see \cite{Remazeilles2020} for details).

\subsection{Component separation method}
\label{subsec:method}

The observed sky data, $d_\nu (\hat{n})$, in the line-of-sight $\hat{n}$ and frequency channel $\nu$ can be written as
\begin{equation}\label{eq:sky}
d_\nu (\hat{n}) = I_\nu^{\rm SZ}(\hat{n}) + n_\nu (\hat{n})\,,
\end{equation}
where $I_\nu^{\rm SZ}(\hat{n})$ is the rSZ signal intensity (Eq.~\ref{eq:rsz}) and $n_\nu (\hat{n})$ is the overall nuisance term including foregrounds and instrumental noise. Following \cite{Chluba2013}, we can perform the Taylor expansion of the rSZ spectral energy distribution (SED) around some pivot temperature $\overline{T}_{\rm e}$ so that
\begin{equation}\label{eq:taylor}
d_\nu (\hat{n}) = f(\nu, \overline{T}_{\rm e})\, y(\hat{n}) + \frac{\partial f(\nu, \overline{T}_{\rm e})}{\partial \overline{T}_{\rm e}}\, \left(T_{\rm e}(\hat{n}) - \overline{T}_{\rm e}\right)y(\hat{n}) + \tilde{n}_\nu (\hat{n})\,,
\end{equation}
where the nuisance term $\tilde{n}_\nu$ now includes higher-order terms in $T_{\rm e}(\hat{n}) $ in addition to foregrounds and noise. The expansion Eq.~\eqref{eq:taylor} thus highlights two distinct components of emission, the Compton parameter component $y(\hat{n})$ and a \emph{temperature-modulated} component ${\left(T_{\rm e}(\hat{n}) - \overline{T}_{\rm e}\right)y(\hat{n})}$, with different spectral signatures $f(\nu, \overline{T}_{\rm e})$ and $\partial f(\nu, \overline{T}_{\rm e}) / \partial \overline{T}_{\rm e}$, respectively. Therefore, it is possible in principle to disentangle the $y$ and ${y(T_{\rm e} - \overline{T}_{\rm e})}$ fields through multi-frequency observations and tailored component separation methods.

Given that both signals $y$ and ${y(T_{\rm e} - \overline{T}_{\rm e})}$ are correlated with each other, it is essential to fully deproject one signal from the estimated map of the other signal, i.e. nulling out $y$ in the reconstructed map of the ${y(T_{\rm e} - \overline{T}_{\rm e})}$ field and vice-versa, by using a \emph{constrained} ILC approach \cite{Remazeilles2011,Remazeilles2020,Remazeilles2021}. Denoting the temperature-modulated Compton-$y$ signal as ${z(\hat{n})\equiv y(\hat{n})\left(T_{\rm e}(\hat{n}) - \overline{T}_{\rm e}\right)}$, we thus form our constrained-ILC estimate $\hat{z}(\hat{n})$ through the following weighted linear combination of the data across frequencies:
\begin{equation}\label{eq:cilc}
\hat{z}(\hat{n}) = \sum_\nu w_\nu\, d_\nu (\hat{n}) \quad \textrm{such that} \quad 
\begin{cases}
\textrm{the variance } \langle\, \hat{z}^{\,2} \rangle \textrm{ is minimum}\,,\\[1.5mm]
\sum_\nu w_\nu\, \frac{\partial f(\nu, \overline{T}_{\rm e})}{\partial \overline{T}_{\rm e}} = 1\,, \\[1.5mm]
\sum_\nu w_\nu\, f(\nu, \overline{T}_{\rm e}) = 0\,.
\end{cases}
\end{equation}
The unit constraint in Eq.~\eqref{eq:cilc} ensures the conservation of the signal of interest, ${z \equiv y(T_{\rm e} - \overline{T}_{\rm e})}$, during variance minimization, while the null constraint guarantees full cancellation of $y$ residuals in the recovered ${y(T_{\rm e} - \overline{T}_{\rm e})}$-map. Finally, the minimum-variance condition in  Eq.~\eqref{eq:cilc} ensures the mitigation of foregrounds and noise. Note that the $y$-map can be obtained in a similar way, simply by interchanging the unit and null constraints in Eq.~\eqref{eq:cilc}. In addition to above constraints in Eq.~\eqref{eq:cilc} we add two extra nulling constraints to deproject the kinetic SZ effect (hence the CMB) and bulk of the thermal dust emission in the recovered rSZ maps (see \cite{Remazeilles2020} for details).
The expression of the constrained-ILC weights $\boldsymbol{w}\equiv\{w_\nu\}$ for our specific component separation problem Eq.~\eqref{eq:cilc} can be derived using Lagrange multipliers (see \cite{Remazeilles2020}):
\begin{equation}\label{eq:filter}
\boldsymbol{w}^{\rm T} = \boldsymbol{e}^{\rm T}\left({\rm A}^{\rm T}{\rm C}^{-1}{\rm A}\right)^{-1}{\rm A}^{\rm T}{\rm C}^{-1}\,,
\end{equation}
where the first two columns of matrix ${\rm A} = \left( \partial \boldsymbol{f}/ \partial \overline{T}_{\rm e}\,\, \boldsymbol{f}\,\, \boldsymbol{g}_{\rm kSZ}\,\, \boldsymbol{g}_{\rm dust}\right)$ collect the effective SED vectors of the rSZ signals (see Eq.~\ref{eq:taylor}) and the last two columns the SED vector $\boldsymbol{g}_{\rm kSZ}$ of the kinetic SZ effect and the average SED vector $\boldsymbol{g}_{\rm dust}$ of the thermal dust, while $\boldsymbol{e}^{\rm T} = \left(1\, 0\, 0 \,0 \right)$, and $C=\langle \boldsymbol{d} \boldsymbol{d}^{\rm T} \rangle$ is the channel-to-channel covariance matrix of the data.

\section{Overview of the results}
\label{sec:results}

\subsection{Electron-temperature spectroscopy of galaxy clusters}
\label{subsec:spectroscopy}

\begin{figure*}
\centering
\includegraphics[width=0.33\columnwidth]{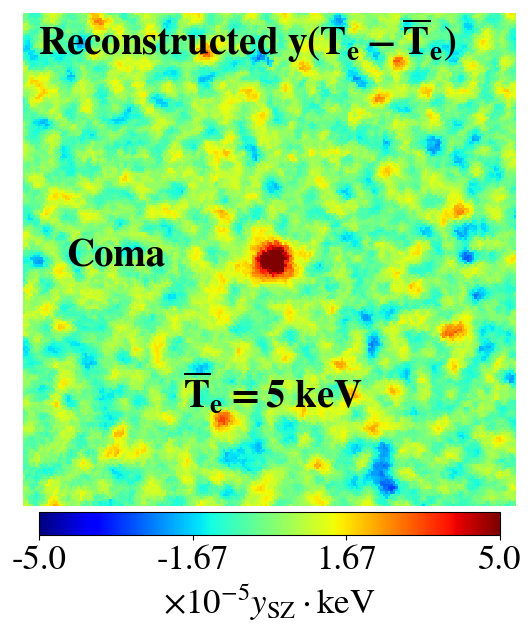}~
\includegraphics[width=0.33\columnwidth]{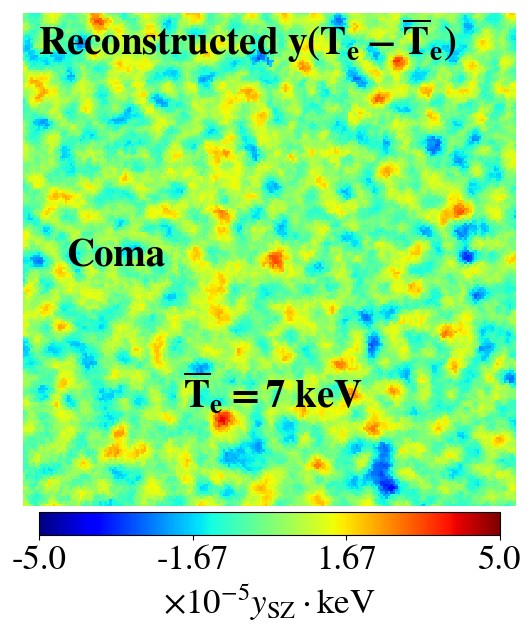}~
\includegraphics[width=0.33\columnwidth]{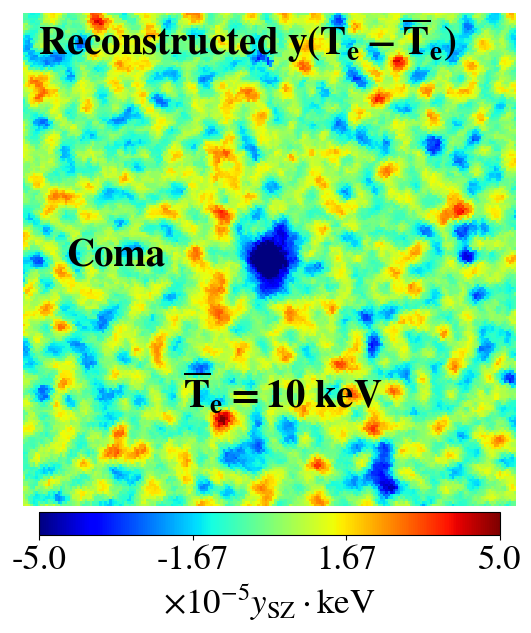}
\caption{Temperature spectroscopy of galaxy clusters: Gnomonic view of the reconstructed $y(T_{\rm e}-\overline{T}_{\rm e})$ all-sky map around Coma for different pivot temperatures $\overline{T}_{\rm e}$.}
\label{fig:spectro}
\end{figure*}

The new rSZ observable ${z(\hat{n})\equiv y(\hat{n})(T_{\rm e}(\hat{n}) - \overline{T}_{\rm e})}$ is particularly interesting to find out the actual temperature $T_{\rm e}(\hat{n})$ of a cluster. Indeed, using different pivot temperatures $\overline{T}_{\rm e}$ for the Constrained-ILC filter Eq.~\eqref{eq:filter} allows us to perform the temperature spectroscopy of the clusters: the recovered signal ${z(\hat{n})\equiv y(\hat{n})(T_{\rm e}(\hat{n}) - \overline{T}_{\rm e})}$ will show a \emph{decrement} at the position of a cluster in the map if the actual temperature of that cluster is smaller than the pivot temperature, i.e. $T_{\rm e}(\hat{n}) < \overline{T}_{\rm e}$, while it will show an \emph{increment} if $T_{\rm e}(\hat{n}) > \overline{T}_{\rm e}$ and a \emph{null} if $T_{\rm e}(\hat{n}) \simeq \overline{T}_{\rm e}$. In the latter case the pivot temperature must be the actual temperature of the cluster.

Such temperature spectroscopy is illustrated in Fig.~\ref{fig:spectro} where we show the reconstructed map of ${y(\hat{n})\left(T_{\rm e}(\hat{n}) - \overline{T}_{\rm e}\right)}$ around the Coma cluster as obtained for different pivots $k\overline{T}_{\rm e}=5$, $7$, and $10$\,keV after component separation on the \emph{PICO} map simulations. The recovered rSZ signal informs us that the temperature of Coma is hotter than $5$\,keV but colder than $10$\,keV, and closer to $7$\,keV (null), which is the fiducial temperature of Coma in the current simulation.

\subsection{Electron gas temperature observables from the rSZ maps}
\label{subsec:te}

From the reconstructed $y$ and $y(T_{\rm e} - \overline{T}_{\rm e})$ maps after component separation, we can easily create the $yT_{\rm e}$-map simply by mutiplying the $y$-map by the pivot temperature $\overline{T}_{\rm e}$ and adding it to the $y(T_{\rm e} - \overline{T}_{\rm e})$-map: ${yT_{\rm e} (\hat{n}) = \overline{T}_{\rm e}\,y(\hat{n}) + z(\hat{n}) = \overline{T}_{\rm e}y(\hat{n}) + y(\hat{n})\left(T_{\rm e}(\hat{n}) - \overline{T}_{\rm e}\right)}$.
Using the $y$-map and the $yT_{\rm e}$-map we can then derive several observables which we outline hereafter. 

\begin{figure*}
\centering
\includegraphics[width=0.33\columnwidth]{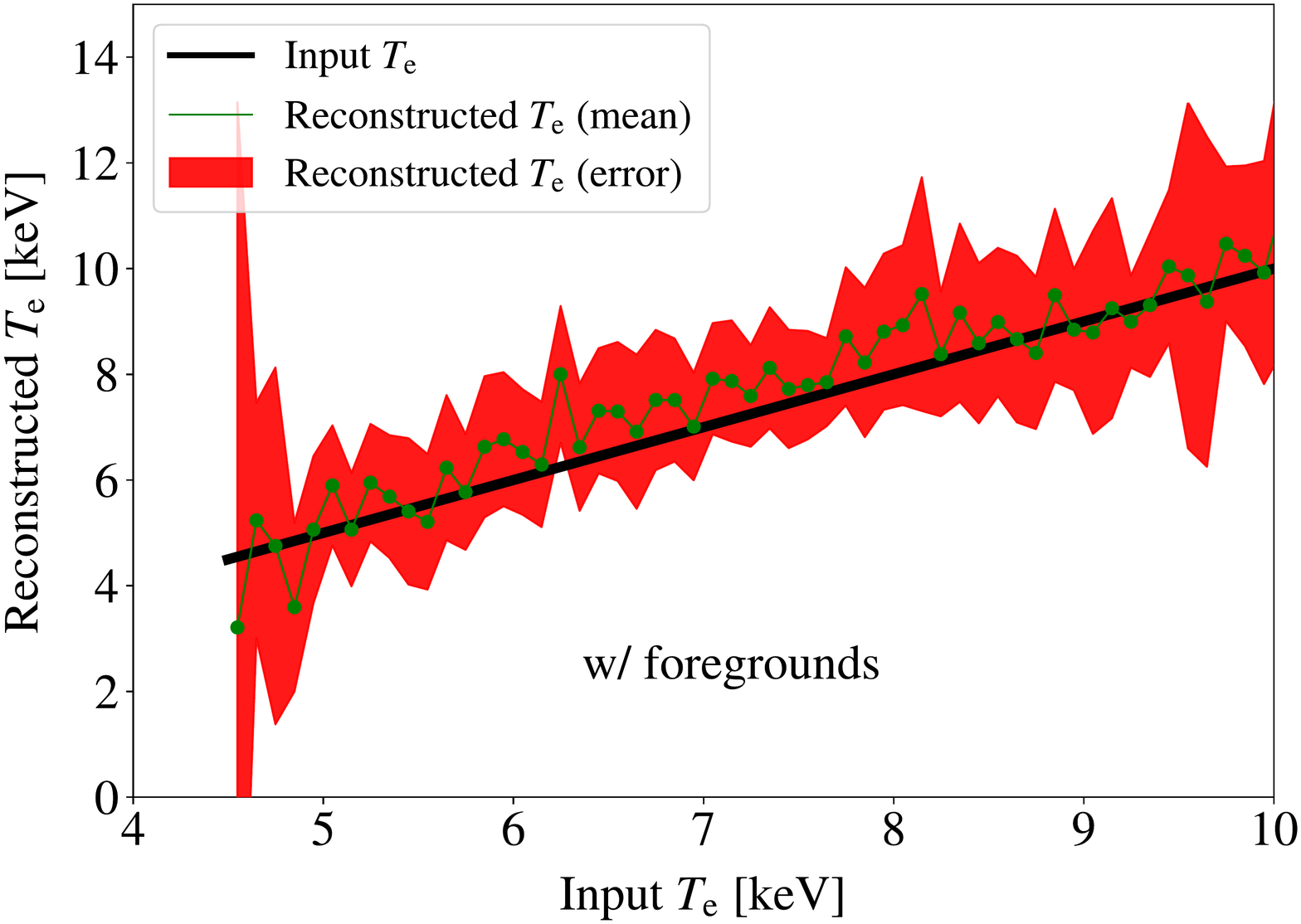}~
\includegraphics[width=0.33\columnwidth]{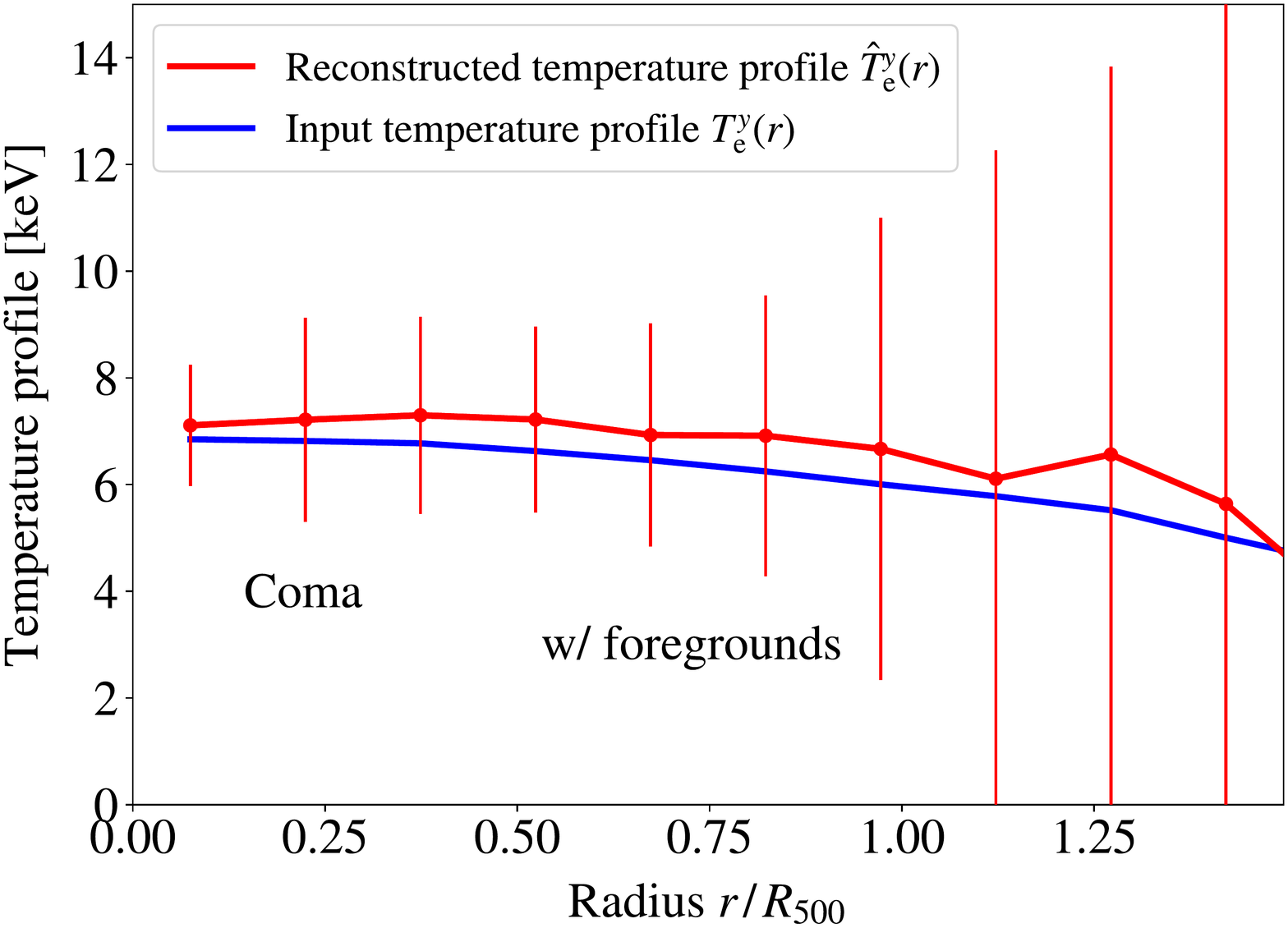}~
\includegraphics[width=0.33\columnwidth]{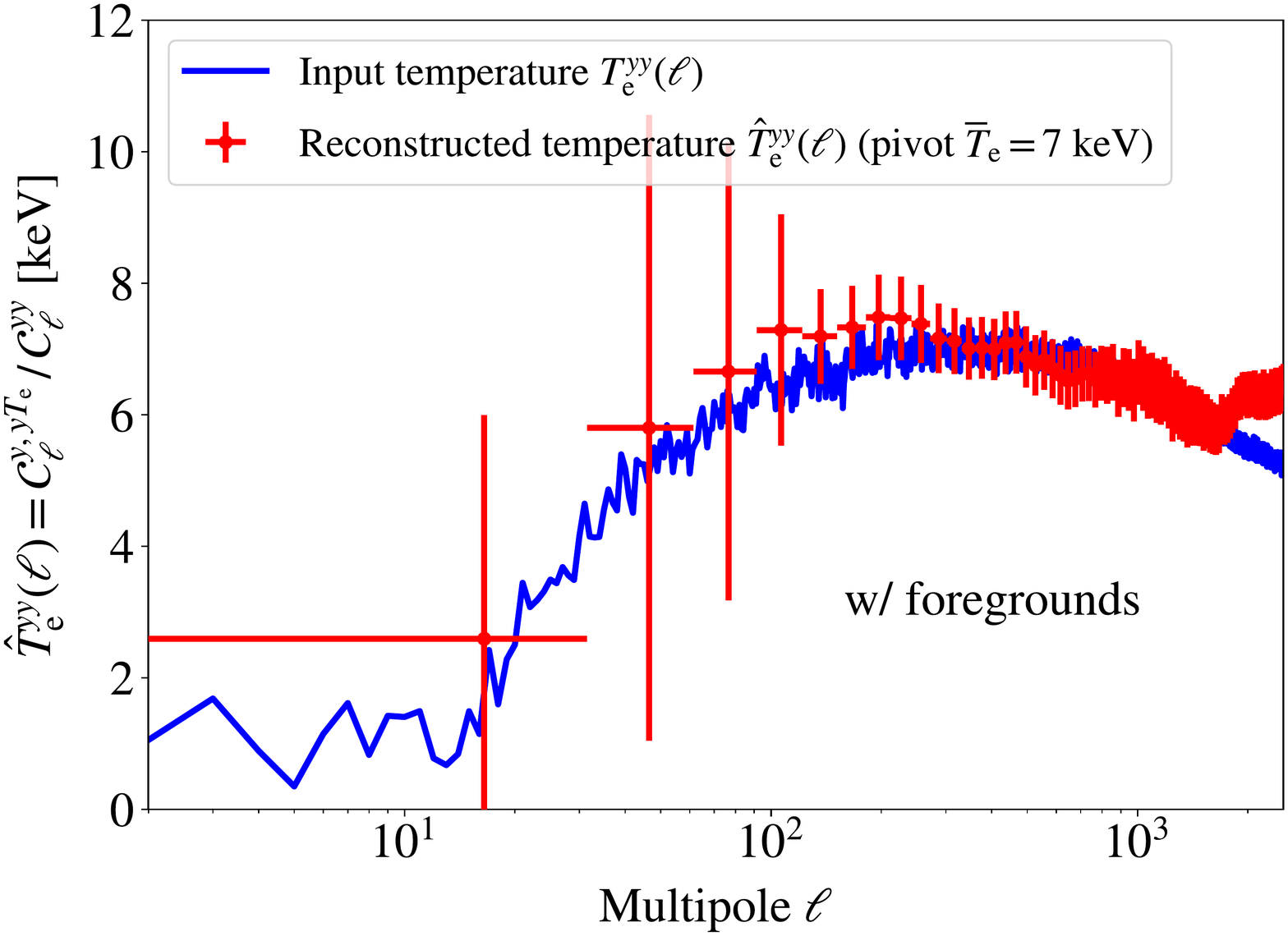}
\caption{\emph{Left}: Recovered electron temperatures $T_{\rm e}[R_{500}]$ across the sky of a sample of 811 clusters after foreground cleaning. \emph{Middle}: Reconstructed temperature profile $T_{\rm e}(r)$ of Coma. \emph{Right}: Reconstructed diffuse electron gas temperature power spectrum $T_{\rm e}^{yy}(\ell)$ after foreground cleaning.}
\label{fig:te}
\end{figure*}

First, by computing the ratio of the flux in the $yT_{\rm e}$-map and the flux in the $y$-map within $R_{500}$ around a cluster, we obtain the $y$-weighted average temperature of that cluster over $R_{500}$:
\begin{equation}\label{eq:tey}
T_{\rm e}^y\left[R_{500}\right] = \frac{ \langle yT_{\rm e}(\hat{n}) \rangle_{\vert \hat{n} - \hat{n}_c\vert \leq R_{500}} }{ \langle y(\hat{n}) \rangle_{\vert \hat{n} - \hat{n}_c\vert \leq R_{500}} }\,,
\end{equation}
where $\hat{n}_c$ denotes the line-of-sight corresponding to the centre of the cluster. 
Figure~\ref{fig:te} (left panel) shows our successful recovery of the electron temperatures of more than 800 galaxy clusters across the entire sky for the \emph{PICO} sky simulation after foreground cleaning and component separation \cite{Remazeilles2020}. The recovered rSZ temperatures could be used in principle as a new proxy for determining cluster masses without relying on X-ray measurements.

Second, we can also derive the electron temperature profile of a cluster simply from the ratio of the cluster profile in the $yT_{\rm e}$-map and the cluster profile in the $y$-map:
\begin{equation}\label{eq:teprofile}
T_{\rm e}^y(r) = \frac{ yT_{\rm e}(r) }{ y(r) }\,,
\end{equation}
where $r$ is the radius from the centre of the cluster. 
Figure~\ref{fig:te} (middle panel) shows as an example our recovery of the electron temperature profile of the Coma cluster up to large radii, with a $10\sigma$ measurement of the average temperature of Coma over $R_{500}$ for \emph{PICO}  \cite{Remazeilles2020}.

Finally, the cross-power spectrum between the reconstructed $y$- and $yT_{\rm e}$-maps relative to the auto-power spectrum of the $y$-map measures the $y^2$-weighted average electron temperature over the full sky across multipoles:
\begin{equation}\label{eq:teyy}
T_{\rm e}^{yy}(\ell) = \frac{ \langle \left(yT_{\rm e}\right)_{\ell m}\, y^*_{\ell m} \rangle }{ \langle y_{\ell m}\, y^*_{\ell m} \rangle } = \frac{ C_\ell^{y,yT_{\rm e}} }{ C_\ell^{yy } }\,,
\end{equation}
thus providing an effective power spectrum of the electron temperature field \cite{Remazeilles2019}. 
Figure~\ref{fig:te} (right panel) shows accurate recovery of the electron temperature power spectrum with \emph{PICO} over a large range of multipoles after foreground cleaning and component separation  \cite{Remazeilles2020}.
\vspace{0.4cm}


\noindent{\bf Concluding remarks.} %
The electron temperature power spectrum $T_{\rm e}^{yy}(\ell)$ provides a new map-based observable, complementing the $y$-map power spectrum $C_\ell^{yy}$, to constrain cosmological parameters, since the exact shapes of the power spectra $T_{\rm e}^{yy}$ and $C_\ell^{yy}$ have different scaling and dependence on cosmological parameters. To gain intuition on this, we may look into theoretical expressions:
\begin{align}
\label{eq:theory1}
C_{\ell}^{yy}  &= \int_{\,0}^{\,z_{\rm max}} dz{dV\over dz}  \int_{\,M_{\rm min}}^{\,M_{\rm max}} dM\, {dn(M,z) \over dM}\,\vert y_\ell (M,z)\vert^2\,,\\
\label{eq:theory2}
C_{\ell}^{y,yT_{\rm e}}  &= \int_{\,0}^{\,z_{\rm max}} dz{dV\over dz}  \int_{\,M_{\rm min}}^{\,M_{\rm max}} dM\, {dn(M,z) \over dM}\,T_{\rm e}(M,z)\,\vert y_\ell (M,z)\vert^2\,.
\end{align}
While $C_{\ell}^{yy}$ and $C_{\ell}^{y,yT_{\rm e}}$ have same scaling with the mass bias $b$ through exact same dependence on the profile $\vert y_\ell (M,z)\vert^2$, they have different scaling with $\sigma_8$ due to the modulation of the halo mass function in $C_{\ell}^{y,yT_{\rm e}}$ by the mass-dependent temperature $T_{\rm e}$. Therefore, while the $y$-map power spectrum $C_{\ell}^{yy}$ depends on $\sigma_8$ and $b$ in a degenerate form, the electron temperature power spectrum $T_{\rm e}^{yy}(\ell)$, as the ratio of $C_{\ell}^{y,yT_{\rm e}}$ and $C_{\ell}^{yy}$ (Eq.~\ref{eq:teyy}), must depend only on $\sigma_8$ and be quite insensitive to the mass bias \cite{Remazeilles2020}. Such new rSZ observable thus provides a possible avenue to break some of the parameter degeneracies with future cosmological SZ analyses.
\newline

\noindent\emph{Acknowledgements}: This work is supported by the ERC Consolidator Grant CMBSPEC (No. 725456) as part of the European Union's Horizon 2020 research and innovation program. 
 
%

\begin{thebibliography}{}
%
%
\bibitem{sz}
Zeldovich Ya. B., Sunyaev R. A., Astrophys. Space Sci., \textbf{4}, 301 (1969)
\bibitem{Chluba2013}
Chluba J., Switzer E.,  Nelson K., Nagai D., MNRAS, \textbf{430}, 3054 (2013)
\bibitem{Planck_sz_2011}
Planck Collaboration, A\&A, \textbf{536}, A8 (2011)
\bibitem{Planck_sz_2015}
Planck Collaboration, A\&A, \textbf{594}, A27 (2016)
\bibitem{Planck_ymap_2015}
Planck Collaboration, A\&A, \textbf{594}, A22 (2016)
\bibitem{Arnaud2005}
Arnaud M., Pointecouteau E., Pratt G. W., A\&A, \textbf{441},  893 (2005)
\bibitem{Hurier2016}
Hurier G., A\&A, \textbf{596}, A61 (2016)
\bibitem{Erler2018}
Erler J., Basu K., Chluba J., Bertoldi F., MNRAS, \textbf{476}, 3360 (2018)
\bibitem{Remazeilles2019}
Remazeilles M., Bolliet B., Rotti A., Chluba J., MNRAS, \textbf{483}, 3459 (2019)
\bibitem{Hazumi2020}
Hazumi M., et al., in SPIE Conference Series. p. 114432F (2020)
\bibitem{Hanany2019}
Hanany S., et al., arXiv:1902.10541 (2019)
\bibitem{Remazeilles2020}
Remazeilles M., Chluba J., MNRAS, \textbf{494}, 5734 (2020)
\bibitem{Remazeilles2011}
Remazeilles M., Delabrouille J., Cardoso J.-F., MNRAS, \textbf{410}, 2481 (2011)
\bibitem{Remazeilles2021}
Remazeilles M., Rotti A., Chluba J., MNRAS, \textbf{503}, 2478 (2021)
\end{thebibliography}
%
%

\end{document}